# The Role of Counterions in the Assembly of Charged Virus-Like Shells

Ya-Wen Hsiao[1]*, Magnus Hedström[2], Maxim G Ryadnov[3,4], David J Bray[1], and Jason Crain[,5,6]

[1] The Hartree Centre, STFC Daresbury Laboratory, Warrington, WA4 4AD, UK

[2] Clay Technology, Ideon Science Park, SE-223 70 Lund, Sweden

[3] National Physical Laboratory, Hampton Road, Teddington, TW11 0LW, UK

[4] Department of Physics, King's College London, Strand Lane, London, WC2R 2LS, UK

[5] IBM Research Europe, Hartree Centre, Daresbury, WA4 4AD, UK

[6] Department of Biochemistry, University of Oxford, Oxford OX1 3QU, UK

**KEYWORDS**

Virus-like particles, Protein-based VLP, Synthetic VLP, MD Simulation, Counterion condensation

**ABSTRACT**

Synthetic virus-like particles (VLP), designed from simplified building blocks, can reduce the complexity of native viral proteins and be tailored for specific applications. Using molecular dynamics simulations, we investigate the role of counterions ($H_2PO_4^-$, $PO_4^{3-}$, $Cl^-$, and $F^-$) in stabilizing a pre-assembled virus-like shell formed by cationic peptides exemplified by synthetic cyclopeptide (sequence: cyc-Gln-DLeu-Arg-DLeu-Arg-DLeu-Arg-DLeu) VLP shells. Our findings reveal that polyatomic anions ($H_2PO_4^-$ and $PO_4^{3-}$) facilitate stable assemblies by condensing on the VLP shell to higher degrees, thereby neutralizing Coulombic repulsion among the peptide building blocks. Cohesion is largely promoted through the multidentate hydrogen bonds with arginine: Substituting arginine by lysine in the system with $H_2PO_4^-$ leads to destabilization of the structure. $H_2PO_4^-$ additionally engages in hydrophobic interactions with leucine side chains. By contrast, monoatomic anions ($Cl^-$ and $F^-$) show insufficient coordination to the peptides and fail to stabilize the assembly, while supplementing $F^-$ with excess 1M NaCl can recover the structural integrity by screening electrostatic interactions. This study provides important insights into the role of counterions in molecular self-assembly and the nature of their interactions with amino-acid side chains involved in the cooperative formation and stabilization of synthetic virus-like shells.

## 1. INTRODUCTION

Viruses present a diverse range of naturally occurring nanomaterials that infect all forms of life including plants, animals, archaea and bacteria. Due to their conserved architectures and biological functions that can be mimicked by design, viruses have inspired the development of synthetic and semi-synthetic virus-like particles (VLPs) aiming at various applications. With intracellular gene delivery, which viruses mastered to perfection, VLPs are explored to capitalize on specific viral properties e.g., the ability to encapsulate molecular cargo applicable to therapy, imaging, and vaccine development.[1,2] Furthermore, like other protein-based nanoparticles, protein-based VLPs have the advantages of suitable sizes, biocompatibility, and biodegradability.

The assembly of VLPs rely on stable protein-protein interactions (PPIs).[3] PPIs comprise both hydrophobic and hydrophilic interactions, and are often specified by amphiphilicity, which helps define the shape and morphology of a given assembly.[4–6] Additionally, protein-based VLPs with charged side chains in their constituent amino acids can interact with counterions, similar to the peptide-based drugs that are frequently formulated as salts.[7] Consequently, the influence of ions on PPIs is a significant factor in the formation and stability of VLPs. Studies have been conducted to explore the impact of salts or buffers on PPIs. For example, the polarizability of anions has been shown to modulate PPIs[8]; the addition of salt to protein solutions can induce oligomerization, aggregation, and precipitation,[8–10] which prove to be characteristic of other aqueous biological and colloidal systems that are subject to similar effects of counterions and excess salt.[11–16]

The assembly of charged VLPs in the presence of counterions is closely related to the well-studied phenomenon of counterion condensation observed with macromolecules. This process has been investigated both experimentally and theoretically.[17–23] Through the binding of oppositely charged ions to the constituent residue side chains on the protein surface, long-range electrostatic interactions are neutralized, enabling the formation of short-range ion-bridging forces that promote protein association.[24,25] However, different ions, even of the same charge, can produce varied effects. For instance, the Hofmeister series categorizes ions based on their ability to influence protein solubility, distinguishing between "salting-in" and "salting-out" behaviors.[11] This series has been widely applied to characterize salt effects on protein solubility, polymer phase transitions, and the solubility of small molecules.[13,26–28] Polyatomic or polydentate anions, which can support several hydrogen bonds, such as phosphate and sulfate exhibit stronger interactions with cationic peptide side chains compared to monovalent anions like chloride, as observed in electrophoretic studies.[21,22] However, all these observations are at the macroscopic scale, prompting the need for atomistic-level investigations to fully understand the mechanisms.

The study by Noble et al.[29] introduced a novel synthetic VLP, which is based on the assembly of cationic cyclopeptides into virus-like shells, dubbed CycVir. The assembly is driven by the hydrophobicity of leucine residues, and electrostatic interactions between arginine residues and phosphate counterions via coordination. The current manuscript aims to contribute insights into the counterion selection in modulating the stability of CycVir. The effects of four distinct counterions, namely $H_2PO_4^-$, $PO_4^{3-}$, $Cl^-$, $F^-$, on the structural stability of CycVir were investigated. These anions were chosen because of their different salting-out propensities according to the Hoffmeister series. In addition, $H_2PO_4^-$ and $PO_4^{3-}$ are main components in the phosphate buffer used as a physiological medium. Trivalent as well as divalent counterions, are well known to enhance cohesion between charged macromolecules by mediating ion-ion correlations and acting as bridges.[30,31] Although the focus of this study is not on such multivalent-mediated binding, the inclusion of $PO_4^{3-}$ allows us to partly address this effect through comparison with other monovalent anions. By understanding the effects of the counterions, this study provides a complementary understanding of the design principles and points to practical parameters for optimizing VLP systems intended for use in different environments.

Many valuable insights into the formation and stability of VLP shells have been obtained from coarse-grained models, particularly those that explore the interplay between electrostatic and hydrophobic interactions.[32–36] However, such models often lack explicit molecular resolution, making it challenging to capture detailed peptide-counterion interactions and hydrogen-bonding patterns. By detailing interactions at atomistic scale obtained from the molecular dynamics (MD) simulations, this study complements these coarse-grained approaches with specific ion effects that influence shell stability.

## 2. METHODS

The CycVir model (Figure 1a), as described in the work of Noble et al.,[29] was adopted. This model is assembled from tessellation units of three stacked peptides (sequence: cyc-Gln-DLeu-Arg-DLeu-Arg-DLeu-Arg-DLeu, charge +3, shown in Figure 1b) totaling 432 cyclopeptides, with an initial radius set to 45 Å. The units were uniformly distributed on a sphere, with the normal of the peptide rings pointing radially. Experimentally, CycVir was found to be polydisperse.[29] We chose to start the simulation at the lower end of the experimentally reported size range taking measurement uncertainty into account, to promote hydrophobic interactions among peptides to establish from the outset. Conceivably, this initial packing could introduce local clashes, which may lead to expansion as the system relaxes toward an optimal structure. Figure 1c shows $\rho_{\text{norm}}(r)$, the radial number density normalized by the number of particles $N$ of the species of interest. Thus, $\rho_{\text{norm}}(r) \cdot N$ gives the local average number density at a distance $r$ from the assembly's center of mass.

MD simulations were performed on this pre-assembled model to observe its stability in the presence of the counterions of choice. Our focus is not on simulating the dynamic processes of assembly or disassembly. This approach is motivated by experimental evidence indicating a low critical aggregation concentration[29], suggesting that once formed, the VLP shells are thermodynamically stable. We then analyze the interactions that contribute to the cohesion for a stable assembly.

The simulations were performed using NAMD 2.12[37] software with the CHARMM36 force field,[38,39] incorporating fluoride parameters from Orabi et al.[40] The model CycVir was solvated in a box (175 Å × 173 Å × 172 Å) with TIP3P water.[41] Periodic boundary conditions were imposed. All simulations were run using Langevin dynamics in the NPT ensemble at 298 K, and 1 bar with an isotropic coupling. The nonbonded interaction cutoff was set to 12 Å and the switch distance to 10 Å. Bonds involving hydrogen were made rigid using the SHAKE algorithm.[42] Electrostatics were calculated using the particle mesh Ewald approach.[43] To balance the net positive charge of the peptides, 1296 $H_2PO_4^-$/$Cl^-$/$F^-$ or 432 $PO_4^{3-}$ ions were randomly placed in the simulation cell. Generally, except in two, the simulations were performed without additional salt.

For the simulations, restrained equilibrium runs totaling over 70 ns were performed prior to the production run. These were carried out in two stages using harmonic restraints on the peptide backbone, first with a force constant of 1 kcal/mol/Å$^2$ and then 0.1 kcal/mol/Å$^2$, both using a time step of 1 fs. The production run used a 2 fs time step, and convergence was considered achieved when the radius of gyration reached a plateau.

To provide a picture of the composition of the converged structure, we examined the distribution of species in the studied systems by calculating $\rho_{\text{norm}}(r)$ for peptides, counterions, and water. These calculations used specific atoms for each species from individual frames, averaged over the final 40 ns, during which the structure had stabilized. The selected atoms were the Cα of the glutamine residue for peptides, the phosphorus atom (P) for polyatomic counterions ($H_2PO_4^-$/ $PO_4^{3-}$), $F^-$, $Cl^-$, and the oxygen atom for water. $\rho_{\text{norm}}(r)$ was evaluated using the center of mass, frame by frame, of the respective species type as the reference point.

Hydrogen bonds (HBs) between species of interest were evaluated for the final structure of each system using the criteria: the distance of 3 Å between two heavy atoms and the minimal angle of 150° formed by the donor, hydrogen, and acceptor atoms. Furthermore, a 3 Å distance was used to determine whether any two particles were associated, such as condensed counterions on peptides. All analyses and visualization were done using VMD[44] scripts.

Additional test systems different from the above setups were constructed to investigate the interactions in more detail:

(1) In general, simulations were performed without additional salt in accordance with the experimental setups. However, to test the effects of Coulomb screening on VLP shell stability, one system referred to as $F^-_{NaCl(1M)}$ was prepared by adding 1M NaCl to the $F^-$ system which in this study was found to be unstable. The 1 M concentration was chosen to provide strong electrostatic screening, corresponding to a Debye length (~3 Å) comparable to a typical contact distance. For the sake of completeness and to test the effect of physiological salinity, an additional system containing 0.15 M NaCl ($F^-_{NaCl(0.15M)}$) was also simulated.

(2) To test whether specific interactions involving Arg play a role in stabilizing the assembly, we made a model referred to as $H_2PO_4^-{}_{R2K}$ by replacing all Arg with Lys in the $H_2PO_4^-$ system that was previously shown to form stable assembly in MD simulations.[29] Furthermore, the choice of using the monovalent $H_2PO_4^-$ system is also inspired by the RNA structure where the phosphate is typically monovalent and sometimes divalent. Such choice makes this test biologically relevant.

## 3. RESULTS AND DISCUSSION

### 3.1. Assemblies Stabilized with $H_2PO_4^-$/ $PO_4^{3-}$ but Dispersed with $F^-$/$Cl^-$

We used $\rho_{\text{norm}}(r)$ of the peptides to determine whether a hollow shell has been formed, as a zero density between the center and a finite radius indicates a hollow structure. The results are illustrated in Figure 2 which reveal that the assemblies are stable with $H_2PO_4^-$ and $PO_4^{3-}$ whereas they are not in the cases of $F^-$ or $Cl^-$. The results in Figure 2 are corroborated by the snapshots of the final configurations for the assemblies in Figure S1. Earlier studies suggest that high net charge leads to slow protein aggregation.[45–51] For example, the amyloid formation by positively charged islet amyloid polypeptide is much slower at low pH than at neutral pH.[52] Similarly here, with a net charge of +3 per peptide unit, the potentially difficult assembly of cationic CycVir peptides seems to be facilitated by the presence of $H_2PO_4^-$ or $PO_4^{3-}$ counterions and results in hollow vesicle structures, as shown in the top two panels of Figure 2. Both assemblies expanded compared to the initial structure, with the radius of gyration stabilizing at 55.5 Å for $H_2PO_4^-$ and 60.9 Å for $PO_4^{3-}$ (Figure S2a). This expansion relieved initial packing clashes, resulting in a thicker and less dense shell without major positional rearrangement. As shown in Figure S2b, the center of mass of each peptide retains the same neighboring relationship as the initial configuration, although the side chains as well as the normal of cyclopeptide rings point in various directions different from the initial highly ordered geometry. Notably, the $PO_4^{3-}$ system exhibited a larger empty volume than the $H_2PO_4^-$ system. This difference in shell size is likely due to distinctions in the counterion charge and their total number, and the interactions of these counterions with the peptides (discussed below).

For the $Cl^-$ system, Figure 2 reveals similar radial distributions for all species, indicating that all peptides as well as $Cl^-$ are solvated by water, resulting in a fully dispersed system. In the case of $F^-$, substantial peptide density is observed near their center of mass and spreads across the entire space. Thus, with F⁻, peptides did not assemble into a hollow particle but instead formed a few oligomers.

$\rho_{\text{norm}}(r)$ of $H_2PO_4^-$ and $PO_4^{3-}$ closely follow those of the peptides, indicating that these counterions condense on the peptide assembly. Conceivably, the extent of counterion condensation correlates with the stability of an otherwise highly charged assembly. Figure 3 shows the number of condensed counterions per peptide. The average numbers of condensed counterions per Arg residue over the last 40 ns of the simulation were 0.9, 0.7, and 0.3 for $H_2PO_4^-$, F⁻, and Cl⁻, respectively. The $PO_4^{3-}$ ions were fully condensed on the peptides. For $PO_4^{3-}$ and $H_2PO_4^-$, the counterions largely neutralize the peptide charge, effectively reducing Coulombic repulsion between peptides. In contrast, F⁻ achieves only partial local charge neutralization, and Cl⁻ provides little reduction, consistent with the dispersed assemblies. Counterions affecting molecular aggregation has been previously reported by Desai et al.[53], who suggested that larger anions form suitable pairs with larger organic cations which is consistent with our results of $H_2PO_4^-/PO_4^{3-}$ and peptides. In the following, we will discuss in detail how $H_2PO_4^-$ and $PO_4^{3-}$ exhibit the high degree of interaction with the peptides.

### 3.2. Details of Interactions in the System

The initial CycVir model is a regular tessellation-like assembly of elementary units assembled from three cyclopeptides stacked together, and the final converged configuration, in case of stable assembly, largely retains the initial neighboring relationship between the units (Figure S2b). Ghadiri et al.[54] reported that cyclic D,L-α-peptides form extended stacks resulting in hollow tubular structures, driven by backbone-backbone hydrogen bonding. Although CycVir by design does not adopt a tubular structure, it is still of interest to assess the degree of peptide association through backbone HBs between peptides within stacks. We calculated the radial distribution function (RDF) of backbone hydrogens around backbone carbonyl oxygens. In both the $PO_4^{3-}$ and $H_2PO_4^-$ systems, the RDF revealed only two such hydrogens within 3.85 Å of each carbonyl oxygen, indicating, if any, weak backbone HB associations. This means that HBs of the backbone are not the main driver of self-assembly, suggesting that additional interactions via peptide side chains are required to form stable assemblies. Furthermore, the lateral interactions of the side chains enable the formation of higher dimensional structures as suggested by Insua et al.[55], who showed that 1D tubular structures can evolve into 2D nanosheets. Below, we report the analyses of HBs, hydration levels, as well as close contacts, and explore how counterion mediates the side chain lateral interactions.

**3.2.1. Hydrogen bonding and hydration analysis:** Electrostatic interactions plays a major role for counterion condensation[22]. Our analysis shows that $PO_4^{3-}$ has the strongest counterion-peptide interaction as indicated by its higher probability of forming HBs (Table 1 and Figure 4). $PO_4^{3-}$ forms an average of six HBs per ion, whereas fewer are formed by $H_2PO_4^-$ and $F^-$. Notably, $Cl^-$ forms no HBs with our calculation criteria, indicating a lack of interaction with peptides which is consistent with the observation that its first hydration shell remains largely unaffected by the presence of peptides, unlike the other anions (Table 2). Table 3 shows the number of counterions that form HBs with two or more peptides thence bridge them. $PO_4^{3-}$ shows a high ratio (94%) of condensed ions forming bridging HBs. The ratios for $F^-$ and $H_2PO_4^-$ are lower and both about 30%. The phosphate systems led to stable assembly while CycVir dispersed in the $F^-$ system. Apart from its strong hydration propensity, the fact that $F^-$ is a monoatomic ion can also be the reason: even though it can coordinate to the Arg of two neighboring peptides, it cannot form multidentate HBs to the same extent as $H_2PO_4^-$ does. Figure 5 depicts counterions forming HBs with multiple peptides: a $PO_4^{3-}$ ion bridging six peptides (a); one $H_2PO_4^-$ ion coordinating with three peptides (b); and one $F^-$ forming connections to two peptides (c). Both $PO_4^{3-}$ and $H_2PO_4^-$ form multidentate connections, while $F^-$ forms monodentate HBs and is conceivably less effective in reducing the Coulomb repulsion between Arg side chains. These findings highlight that the stability of peptide assemblies is not solely determined by the number of hydrogen bonds but also by their multidentate nature that enhances bridging efficiency.

Counterion hydration varies with ion type which may influence peptide assembly stability. Table 2 shows that $PO_4^{3-}$ and $H_2PO_4^-$ exhibit reduced hydration shells in the presence of peptides compared to bulk solution, indicating partial dehydration as they bind to peptides. In contrast, $F^-$ and $Cl^-$ remain largely hydrated, with $Cl^-$ showing negligible interaction with peptides as previously discussed. It is consistent with the HB analysis above, that $F^-$ remains highly hydrated therefore exhibits lower binding affinity to the peptides relative to $PO_4^{3-}$ and $H_2PO_4^-$.

Shifting the focus to the hydration of the peptides themselves: The association of charged peptides, facilitated by the replacement of water in their solvation shells with other peptides, can be driven by counterions, as suggested by the Hofmeister series. We evaluate the average number of water molecules surrounding each peptide: 49 for $PO_4^{3-}$, 33 for $H_2PO_4^-$, 51 for $F^-$, and 57 for $Cl^-$ ion systems, respectively. These results confirm that the properties of the protein–solvent system vary with counterion. Peptides in the $H_2PO_4^-$ system have the lowest hydration levels and are thus mostly precipitated which aligns with the observation of stable assembly. In contrast, peptides in the $Cl^-$ system are the most hydrated and exhibit the least precipitation, consistent with the instability of their assembly. Considering only the monovalent counterions ($H_2PO_4^-$, $F^-$, $Cl^-$), the hydration level of peptides appears to inversely correlate with the stability of the peptide assembly.

However, there are more water molecules per peptide with $PO_4^{3-}$ than with $H_2PO_4^-$ and yet both assemblies are stable. This can be explained by the smaller number of $PO_4^{3-}$ ions (432) compared to $H_2PO_4^-$ ions (1296), resulting in less water being displaced. The stability of the assembly in the case of $PO_4^{3-}$ is likely driven by this multidentate ion's strong tendency to form numerous hydrogen bonds, which matches the ability of Arg side chains to support up to five hydrogen bonds per residue.

**3.2.2. Hydrophobic interactions:** Although $H_2PO_4^-$ forms slightly more hydrogen bonds to the peptides than $F^-$, the sum of hydrogen bonds and coordinated water still falls below the bulk coordination number. This suggests the potential involvement of other interactions between CycVir and $H_2PO_4^-$. To explore this further, interactions between counterions and peptide side chains are examined in more detail. Figure 7 shows the number of peptides whose Arg (a) and D-Leu (b) side chains form contacts with $F^-$, $H_2PO_4^-$, and $PO_4^{3-}$. Notably, $H_2PO_4^-$ establishes significant contacts with not only hydrophilic Arg, but also hydrophobic D-Leu side chains with 420 out of 432 peptides on average, whereas $F^-$ mainly associates with Arg, and $PO_4^{3-}$ exclusively interacts with Arg. This is consistent with the expectation that Arg can form ion pairs with anionic counterions; however, it is striking that $H_2PO_4^-$ also interacts substantially with D-Leu. Conceivably, Arg–phosphate interactions immobilize $H_2PO_4^-$, enabling its dihydrogen moiety to form stable contacts with the side chains of D-Leu residues. Figure S3 provides a representative snapshot illustrating a typical contact to D-Leu by the dihydrogen region of an $H_2PO_4^-$, which incidentally bridges to an Arg of a neighboring peptide. The analysis showed that 70% of the condensed $H_2PO_4^-$ formed bridges between Arg/Gln and D-Leu. Thus, beyond electrostatic interactions with hydrophilic residues, $H_2PO_4^-$ contributes additional stability to the assembly through hydrophobic interactions with D-Leu side chains. This specific binding via hydrophobic interactions differentiates $H_2PO_4^-$ from $F^-$ in CycVir assembly. Lund et al.[56] similarly observed with a model macromolecule that specific anion binding is influenced by local interactions, following two molecular mechanisms: smaller ions like $F^-$ bind through strong, localized charge-charge interactions, while larger anions can associate through a combination of ion pairing and delocalized hydrophobic interactions. Our results demonstrate that counterions along with peptides participate in the interplay of hydrophilic and hydrophobic interactions that sustain the stability of the peptide assembly.

Additional simulation replicas were performed for the $H_2PO_4^-$ and $PO_4^{3-}$ systems to further confirm the observed stability and interaction patterns. Comparisons between the two trajectories revealed excellent consistency, as shown in Supporting Information Figures S4–S10 as well as Tables S1–S3.

**3.2.3. Electrostatic screening with excess salt:** Both ion binding and electrostatic screening impact protein electrostatic interactions.[57–59] Having discussed the former, we proceed to examine the effect of increasing electrostatic screening on an otherwise unsuccessful assembly. Specifically, and as proof of principle, we added 1M NaCl to the $F^-$ system (denoted by $F^-_{NaCl(1M)}$) to test whether enhanced electrostatic screening could stabilize the CycVir assembly. It was found that the shell remained stable (Figure 6a). The radial peptide density shows a hollow vesicle profile with increased magnitude at the region corresponding to the shell radius (Figure 6b). The addition of 1M NaCl, however, neither altered the amount of HBs between $F^-$ and peptides nor changed the amount of $F^-$ condensation (maintaining ~2 ions per peptide). This is consistent with these interactions occurring at distances shorter than the Debye length (~3Å for 1 M NaCl(aq)) and thus not significantly screened by excess salt. In contrast, the inter-peptide Coulomb repulsion, which acts over longer distances, is reduced, allowing, e.g., (1) shorter distance between the side chains of Arg without clearly more F⁻ ions to bridge peptides (Table 3); we see that the coordination number to water remains virtually unchanged (Table 2). (2) hydrophobic interactions among peptides (via D-Leu, as illustrated in Figure 1D in the work of Noble et al.[29]) to stabilize the assembly. We found that each D-Leu side chain, on average, forms at least one contact with a D-Leu side chain of another peptide, as observed in systems with $H_2PO_4^-$ or $PO_4^{3-}$. Thus, we propose that hydrophobicity plays an enhanced role in aggregation in the scenario of $F^-_{NaCl(1M)}$. As for the additional Cl⁻, its $\rho_{norm}(r)$ follows the same distribution as water, again indicating that Cl⁻ does not interact with the peptides.

A natural question is whether stabilization could also occur under physiological conditions of 0.15 M NaCl. To answer this question, we reduced the NaCl concentration of the above model to 0.15 M and performed the simulation of the new model, $F^-_{NaCl(0.15M)}$. Instead of giving a plateaued radius of gyration, the initially packed shell started to break into less connected fragments (Figure S11a). In both the salt-free and 1 M NaCl systems, roughly one-third of the F⁻ ions remain uncondensed (Table 3) and contribute to bulk electrostatic screening. Given the size of the simulation cell, the concentration of free F⁻ is just slightly below physiological salinity. While this additional contribution is negligible in the 1 M NaCl system, it increases the effective ionic strength in the 0.15 M case. Still, the added salt is insufficient to stabilize the shell. In a real, larger system, the relative concentration of free F⁻ would be much lower, implying that the combined (0.15M NaCl + free ions) screening would still be insufficient. The results on solvation, condensation, and hydrogen bonding of fluoride corresponding to $F^-_{NaCl(0.15M)}$, shown in Figure 3 and Tables 1-3, display similarities to those of F⁻ and $F^-_{NaCl(1M)}$. This observation further strengthens that, instead of the above factors, adequate Coulomb screening, as in the $F^-_{NaCl(1M)}$ system, is a prerequisite in order to allow the hydrophobic interaction between peptides in the case of F$^-$ counterions.

**3.2.4. Arginine-phosphate interactions**: Although most of the anions in this study form HBs mainly with Arg, the peptide binding appears more effective for $H_2PO_4^-$ and $PO_4^{3-}$ than F$^-$. Using ionic charges to estimate electrostatics does not explain this difference. Lenton et al. pointed out the impact of Arg-phosphate interactions on the reentrant condensation of proteins.[60] Thus, the specific interactions between Arg and phosphate counterions may bring additional attraction that contributes to the stability of CycVir. The side chain of Arg consists of a guanidinium moiety, whose interactions with phosphate, sulfate, and DNA have also been observed in many previous works.[21,56,61–63]

To test this assumption, we substituted all Arg to Lys in the CycVir–$H_2PO_4^-$ system (denoted by $H_2PO_4^-{}_{R2K}$) since this system is stable in its non-mutated form. Substituting Arg with Lys maintains the charge but removes the guanidinium-phosphate interaction. The substitution results in a disassembled structure halfway through the equilibration process (Figure S11b) and only half of the HBs with peptides being formed compared to that with the non-mutated shell (Table 1). Water coordination numbers of the first hydration shell of $H_2PO_4^-$ differ less in the presence of mutated CycVir (Table 2). Fewer $H_2PO_4^-$ are condensed to the peptides and even fewer of them form bridging HBs (Table 3). This test shows that beyond electrostatic attraction between counter ions and charged residues, the additional contribution of specific interaction (preference of multidentate guanidinium-phosphate interaction in this case) is needed for a stable assembly, consistent with the observed effectiveness of phosphate binding to CycVir. The preference of Arg over Lys is observed in nature: e.g., DNA compaction via Arg-rich protamines,[63] Arg-rich motifs on proteins are known to bind RNA and are involved in regulating RNA processing in viruses and cells,[64] and it has also been shown that the interaction between Arg−phosphate is considerably stronger than that of Lys−phosphate.[60,61,63]

Our study provides insights into the assembly mechanisms and stability of cationic virus-like shells exemplified by CycVir. Namely, counterions (1) modify PPIs resulting in reduced Coulomb repulsion; (2) form hydrogen bonds that bridge peptides; (3) can establish hydrophobic interactions. We also found that specific interactions in the form of multidentate HBs between phosphate and Arg stabilize the peptide assembly. Notably, both phosphate buffer species, $H_2PO_4^-$ and $PO_4^{3-}$ stabilize VLPs albeit through slightly different interacting mechanisms with the peptides. The observed differences between $H_2PO_4^-$, $PO_4^{3-}$ and $F^-$, $Cl^-$ agree with the Hofmeister series[65] and may also offer insights into the origins of the series.

We started the simulations from a pre-assembled structure which may restrict exploration of alternative morphologies. Nevertheless, the observed interactions between residues on peptides and counterions are general and not specific to a particular morphology. These interactions are also relevant to assemblies formed from any charged peptides or molecules, not limited to cyclopeptides. It is also worth noting that each simulation provides robust sampling due to the collective dynamics of 432 peptides and 1296 monovalent counterions, effectively capturing hundreds of thousands of peptide–peptide and peptide–ion interactions. This extensive internal sampling lends statistical weight to the observed behavior, even within a single trajectory. The consistency between independent replicas for the $H_2PO_4^-$ and $PO_4^{3-}$ systems, as documented in Supporting Information Figures S4–S10 and Tables S1–S3, reflects this statistical robustness.

## 4. CONCLUSION

Using MD simulations on the model of CycVir, we demonstrated with atomistic detail that $H_2PO_4^-$ and $PO_4^{3-}$ counterions in contrast to $Cl^-$ and $F^-$, are effective in stabilizing the assembly of cationic virus-like shells by altering the properties of the protein−protein and protein−solvent system, by specific binding, or by locally neutralizing electrostatic interactions through counterion condensation.

Specifically, our results showed that Coulomb interactions between peptides and counterions alone do not fully explain different outcomes by different counterions. The number density profiles of polyatomic anions show higher degrees of counterion condensation on CycVir, achieved by extensive hydrogen bonding for $PO_4^{3-}$, and by hydrogen bonding as well as hydrophobic interaction with the D-Leu side chains for $H_2PO_4^-$. Notably, these interactions bridge peptides and thereby stabilize their assembly. Conversely, $Cl^-$ forms no hydrogen bonds, and although $F^-$ shows peptide binding, it does not bridge peptides extensively nor form hydrophobic interaction with them as does $H_2PO_4^-$ and is therefore insufficient to sustain a stable assembly.

Although high degree of counterion-peptide binding correlates to VLP structure stability by neutralizing the repulsion between the peptides, we showed that the hydrophobic PPI can also sustain the CycVir assembly when the Coulombic repulsion is screened by 1M NaCl demonstrated with the $F^-$ system.

Specific interactions between arginine and phosphate contribute to the effective multidentate peptide binding and in turn the VLP stability, as was shown by substituting arginine with lysine, resulting in dispersed assembly, thereby supporting the importance of arginine-phosphate interaction.

These findings offer broader implications, particularly for designing VLPs with tailored functionalities, e.g., in gene therapies where peptide-phosphate interactions play a crucial role in RNA/DNA encapsulation and release.[66–71]

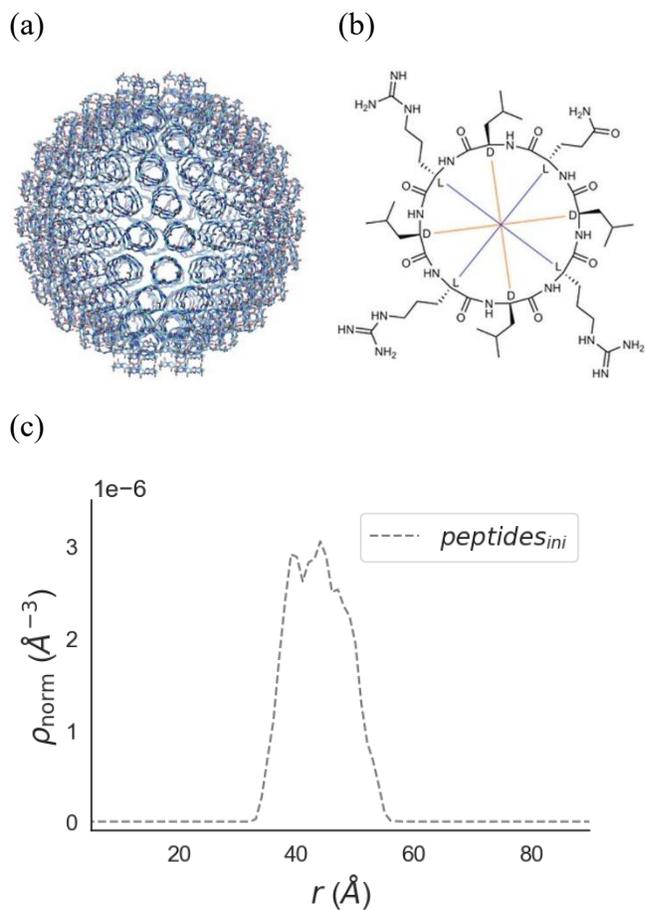

**Figure 1.** (a) Initial configuration of the CycVir system consisting of 432 cyclic peptides, pre-packed into a sphere. (b) Building unit cyclopeptide of CycVir: The figure was reproduced from Figure 1 in ref (29). Copyright 2024 American Chemical Society. (c) The corresponding normalized radial number density $\rho_{\text{norm}}(r)$.

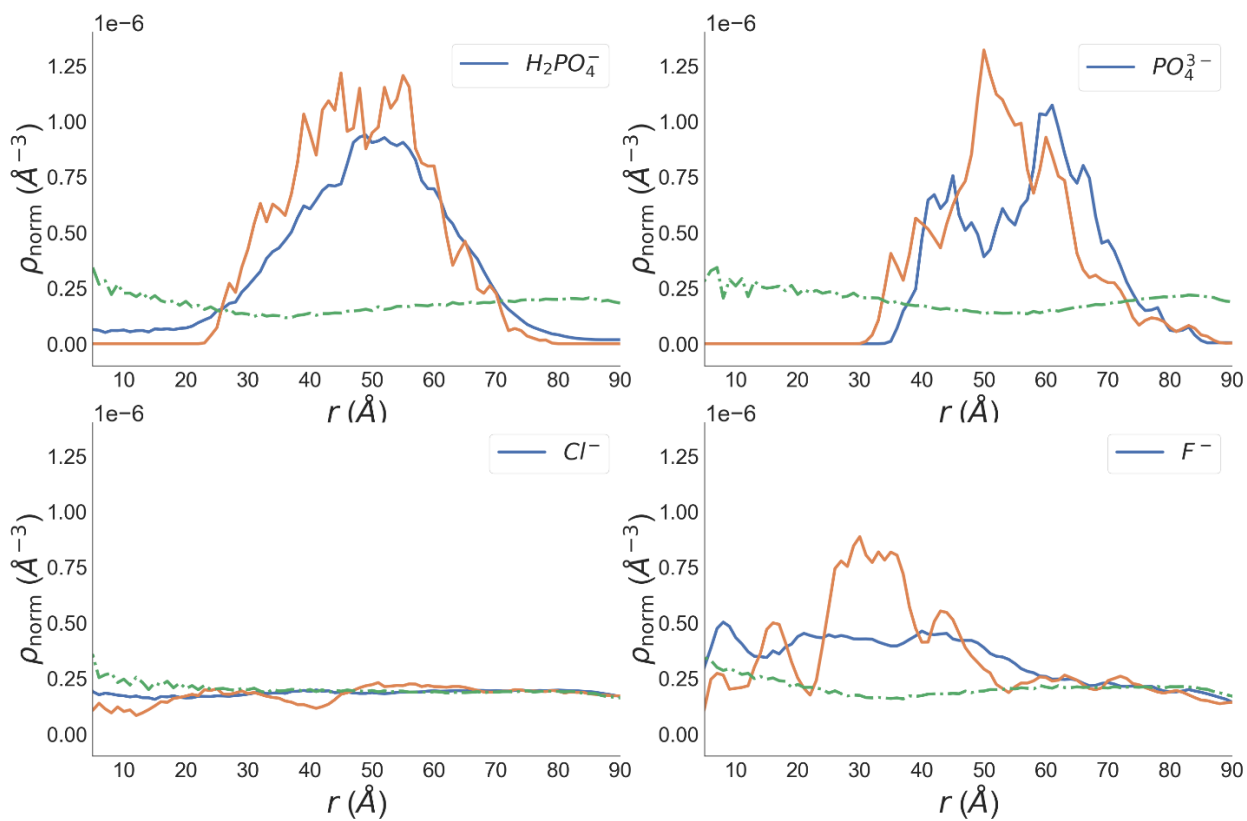

**Figure 2.** Normalized radial number densities $\rho_{\mathrm{norm}}(r)$ of counterions (blue), peptides (orange), and water (green).

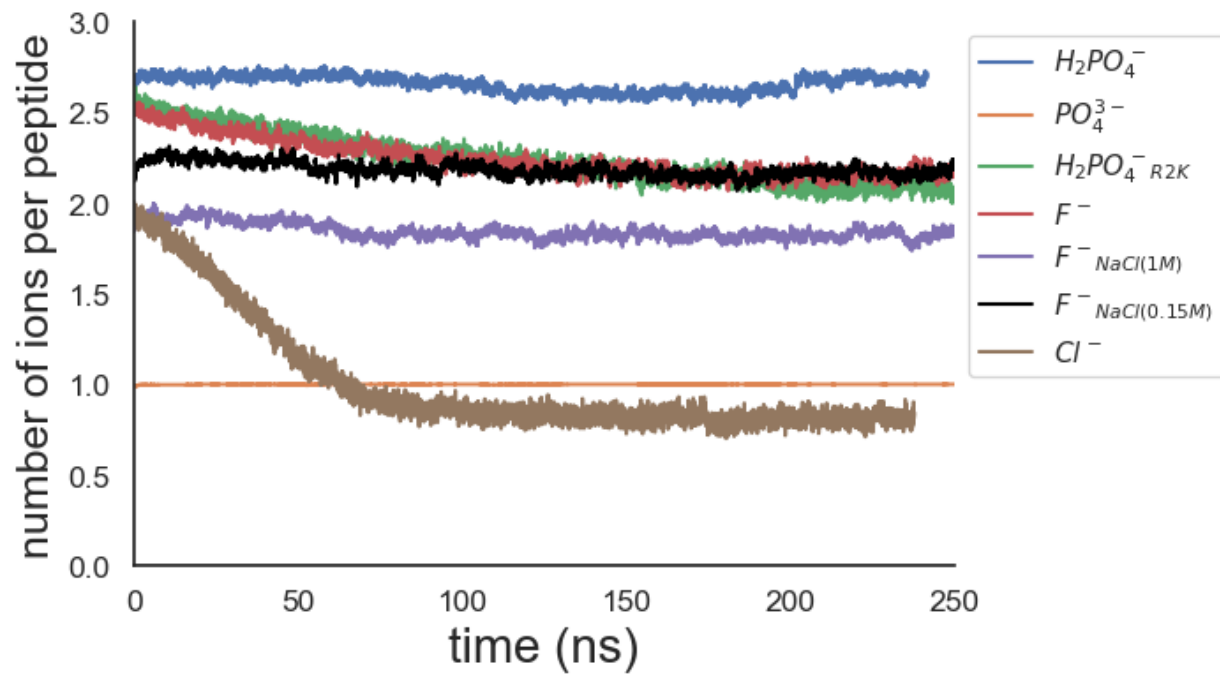

**Figure 3.** Time evolution of the number of condensed counterions per peptide.

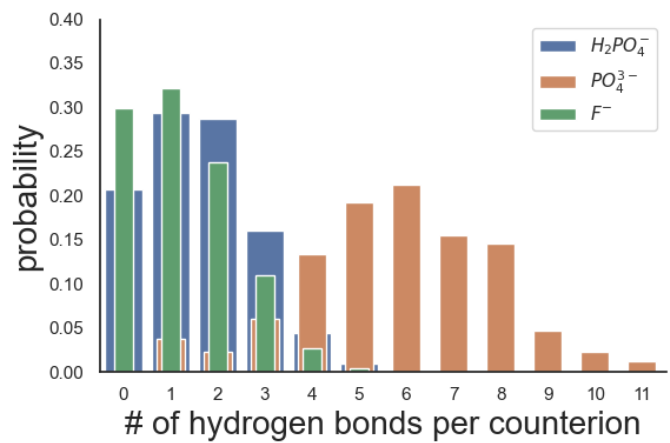

**Figure 4.** Normalized probabilities (per anion) of the number of hydrogen bonds formed between each counterion and peptides.

(a)

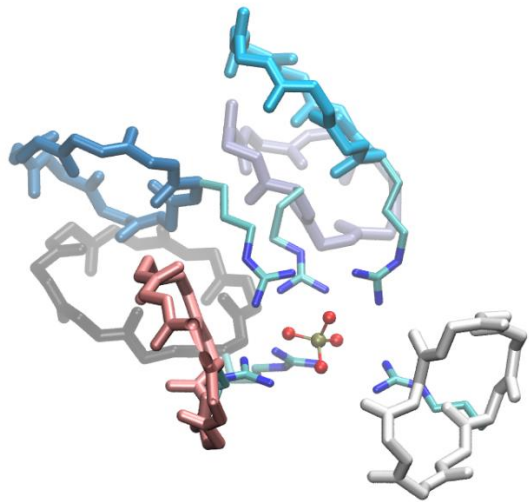

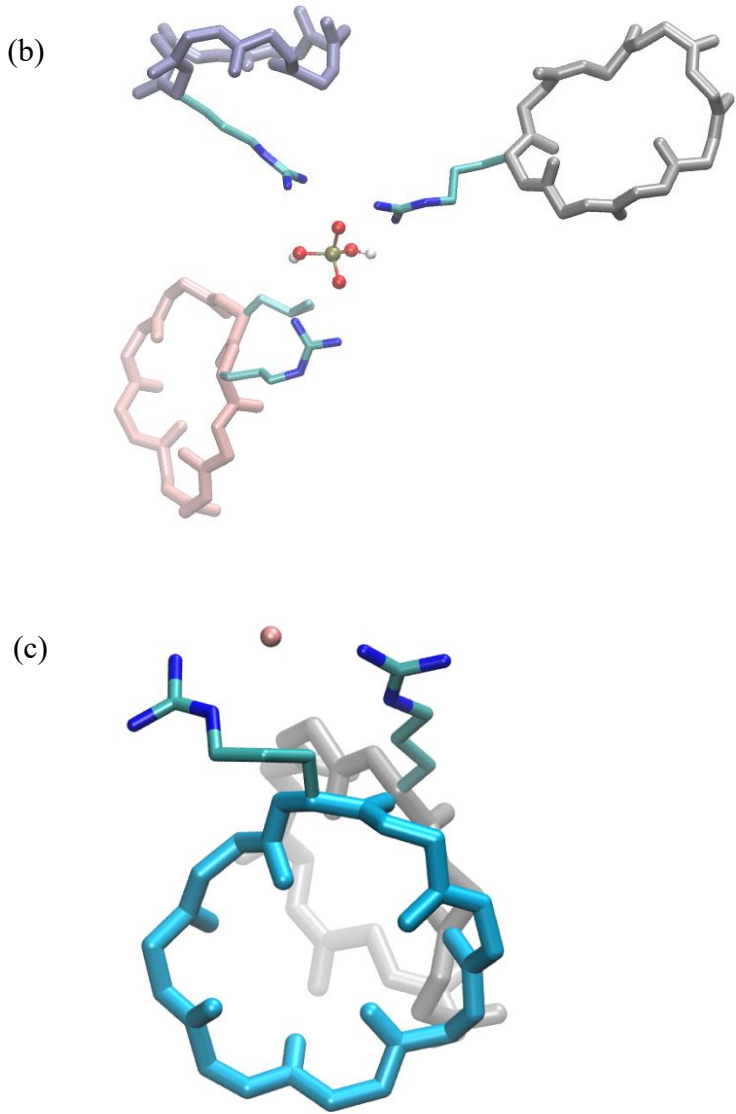

**Figure 5.** Representative snapshots of the various counterions forming HBs with Arg residues of neighboring peptides: (a) one $PO_4^{3-}$ connecting to six peptides, (b) one $H_2PO_4^-$ bridging three peptides, with an additional hydrophobic interaction with D-Leu, and (c) one $F^-$ coordinating to two peptides. For clarity, only the interacting Arg/D-Leu side chains are shown.

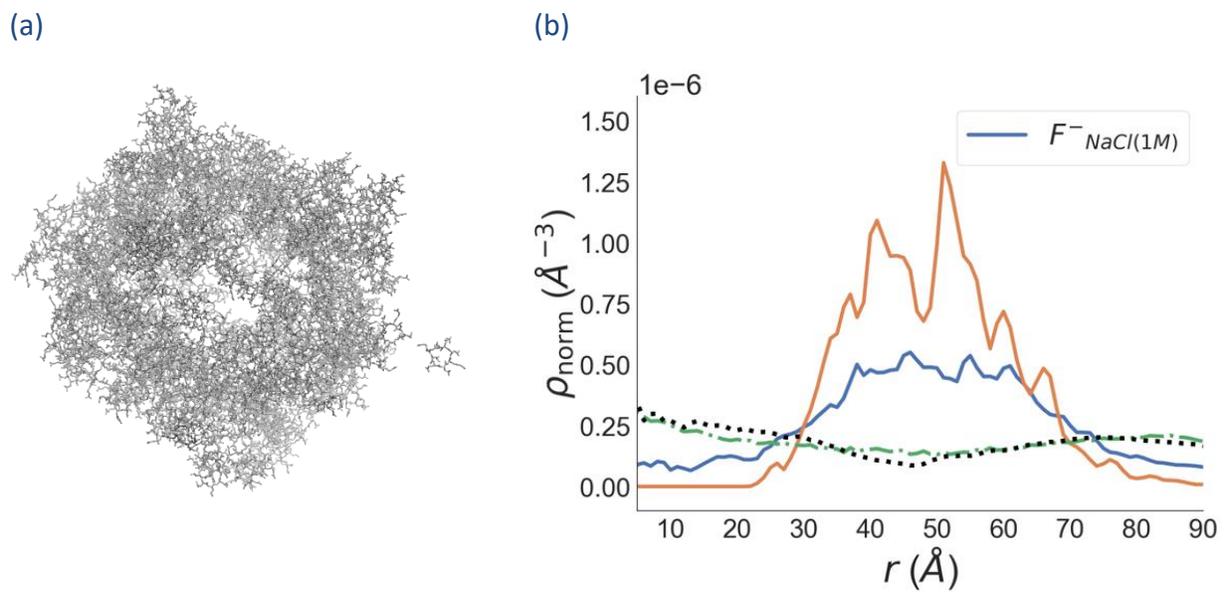

**Figure 6.** (a) Snapshots at the end of the simulations of the stable CycVir shell with $F^-_{NaCl(1M)}$. (b) Normalized radial number densities $\rho_{norm}(r)$ of $F^-$ (blue), peptides (orange), and water (green); with 1M NaCl ($Cl^-$ in black).

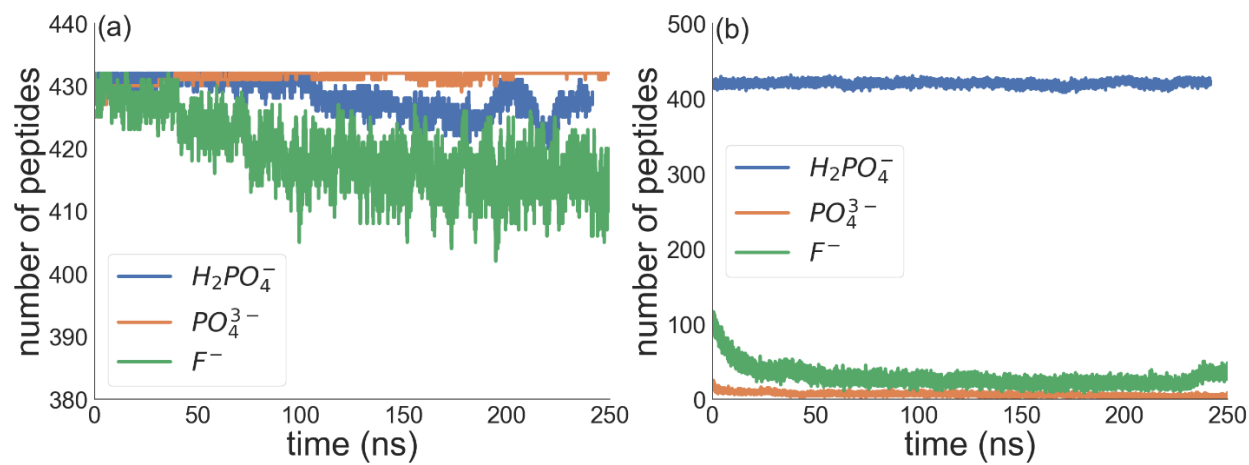

**Figure 7.** Number of peptides whose side chains of Arg (a) D-Leu (b) are in contact with counterions.

**Table 1.** Average number (rounded to the nearest integer) of hydrogen bonds between each counterion and peptides

| Ion | Number of HBs |
|---|---|
| $Cl^-$ | 0 |
| $F^-$ | 1 (100% to Arg) |
| $F^-_{NaCl(1M)}$ | 1 (92% to Arg) |
| $F^-_{NaCl(0.15M)}$ | 1 (91% to Arg) |
| $H_2PO_4^-$ | 2 (81% to Arg) |
| $H_2PO_4^-{}_{R2K}$ | 1 (71% to Lys) |
| $PO_4^{3-}$ | 6 (100% to Arg) |

**Table 2.** Water coordination numbers of the first hydration shell to the anions in bulk solution and in the presence of peptides

| Ion | Number of water oxygens | | Distance of the 1$^{st}$ minimum from P/F/Cl | |
|---|---|---|---|---|
| | **Bulk** | **With peptides** | **Bulk** | **With peptides** |
| $Cl^-$ | 7.7 | 7.3 | 3.75 Å | 3.85 Å |
| $F^-$ | 7 | 5 | 3.35 Å | 3.35 Å |
| $F^-_{NaCl(1M)}$ | 6.8 | 5 | 3.35 | 3.35 Å |
| $F^-_{NaCl(0.15M)}$ | 6.8 | 5 | 3.35 | 3.35 Å |
| $H_2PO_4^-$ | 16 | 8 | 4.95 Å | 4.95 Å |
| $H_2PO_4^-{}_{R2K}$ | 16 | 12 | 4.95 Å | 4.95 Å |
| $PO_4^{3-}$ | 14 | 9 | 4.35 Å | 4.45 Å |

**Table 3.** Number of counterions condensed to the peptides, and number of counterions form HBs with side chains of Arg and Gln from two or more peptides. Also shown are bridging ratios evaluated with respect to the total number of counterions (ratio$_{total}$), as well as that evaluated with respect to the number of counterions condensed to the peptides (ratio$_{condensed}$). Results were based on the final structure of the respective system.

| Ion | Total | Ions condensed to the peptides | Ions bridging the peptides | Bridging ratio$_{total}$ | Bridging ratio$_{condensed}$ |
|---|---|---|---|---|---|
| $F^-$ | 1296 | 941 | 271 | 21% | 29% |
| $F^-_{NaCl(1M)}$ | 1296 | 808 | 296 | 23% | 37% |
| $F^-_{NaCl(0.15M)}$ | 1296 | 934 | 343 | 26% | 37% |
| $H_2PO_4^-$ | 1296 | 1165 | 372 | 29% | 32% |
| $H_2PO_4^-{}_{R2K}$ | 1296 | 918 | 70 | 5% | 8% |
| $PO_4^{3-}$ | 432 | 432 | 406 | 94% | 94% |

**Supporting Information**.

Figures as described in the text including snapshots of final configurations of VLP assemblies; radii of gyration of all systems, and superposition of the center of mass of each peptide in the $H_2PO_4^-$ system; snapshot of $H_2PO_4^-$ forming polar contact with D-Leu; Snapshot of the dispersed mutated CycVir (R->K) in the presence of $H_2PO_4^-$ counterions; Snapshot of the dispersed of CycVir with $F^-$ as counterion and 0.15M extra NaCl; results for 2 replicas of $H_2PO_4^-$ and $PO_4^{3-}$ (PDF)

AUTHOR INFORMATION

**Corresponding Author**

*Ya-Wen Hsiao

The Hartree Centre, STFC Daresbury Laboratory, Warrington, WA4 4AD, UK

E-mail: ya-wen.hsiao@stfc.ac.uk

Tel: +44 (0)1925 603190

**Author Contributions**

Conceptualization, Y.-W.H. and M.H.; Investigation, Y.-W.H.; Data analysis, Y.-W.H. and M.H.; Experimental insights on VLPs, M.G.R.; Funding acquisition, D.B. and J.C.; Writing – original draft, Y.-W.H. and M.H.; Writing – review and editing, Y.-W.H., M.H., M.G.R., D.B., and J.C.

**Funding Sources**

This work was funded by the Hartree National Centre for Digital Innovation (HNCDI) from UK Research and Innovation.

**Notes**

The authors declare no competing financial interest.

**ACKNOWLEDGMENTS**

This work was funded by the Hartree National Centre for Digital Innovation (HNCDI), a collaboration between STFC and IBM.

ABBREVIATIONS

VLP, virus-like particles; PPI, protein-protein interaction.

SYNOPSIS

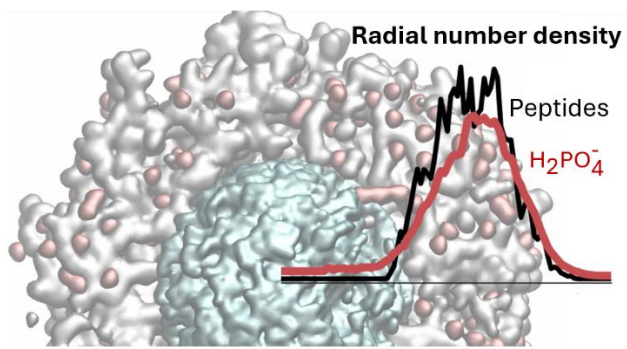